\documentclass[astron]{eas}
\usepackage{graphicx}
\usepackage{natbib}
\usepackage{url}
\newcommand{\msun}{M$_{\odot}$}

\begin{document}

\title{Fossil magnetic fields \\ in intermediate-mass and massive stars} 
\runningtitle{Fossil magnetic fields in intermediate-mass and massive stars}
\author{E. Alecian}\address{Univ. Grenoble Alpes, CNRS, IPAG, F-38000 Grenoble, France}
\author{F. Villebrun$^1$}
\author{J. Grunhut}\address{Dunlap Institute for Astronomy and Astrophysics, Univ. of Toronto, ON M5S 3H4, Canada}
\author{G. Hussain}\address{European Southern Observatory, D-85748 Garching bei M\"unchen, Germany}
\author{C. Neiner}\address{LESIA, Obs. de Paris, PSL Research Univ., CNRS, Sorbonne Univ., UPMC Univ. Paris~06, Univ. Paris Diderot, Sorbonne Paris Cit\'e, F-92195 Meudon, France}
\author{G.~A.~Wade}\address{Department of Physics, Royal Military College of Canada, Kingston, ON, K7K 7B4, Canada}
\author{The BinaMIcS collaboration}\address{http://binamics.lesia.obspm.fr/}
\begin{abstract}
A small fraction of the population of intermediate-mass and massive stars host strong and stable magnetic fields organised on large scales. These fields are believed to be remnants of star formation. It is however not clear how such fossil fields have been shaped during their formation and subsequent evolution. We report recent and ongoing studies on the magnetic properties of pre-main sequence stars and main sequence binaries, allowing us to make progress in this field.
\end{abstract}
\maketitle
\section{Our current knowledge of magnetic fields in intermediate- and high-mass stars}

Stellar fossil magnetic fields, as defined by \citet{borra1982}, are inherent magnetic fields in radiative interiors, i.e. they appear not to be continuously maintained against ohmic decay by ongoing dynamo processes. Until recently, these fields were unambiguously and predominantly found at the surfaces of A- and B-type main sequence (MS) stars of intermediate masses (from $\sim$1.5 to $\sim$8 \msun), but only in 5 to 10\% of them. They display dominant low-order topologies - mainly dipolar and sometimes quadrupolar or octupolar - or a superimposition of low-order fields. The dominant fields are strong, relative to the dipolar field of the Sun, with polar strengths ranging from $\sim$300~G up to 30~kG, peaking in incidence around a few kG. Fossil fields weaker than 300~G appear to be essentially absent in A/B stars \citep{auriere2007}. Below a few G, magnetic fields of different properties have been detected in a few stars, and probably have a different nature \citep[e.g.][]{petit2011,blazere2016a,blazere2016b}. This defines the "magnetic dichotomy'' of the A/B stars. In addition, fossil fields are observed to be stable, showing no detectable variations of topology or strength on timescales of years or decades. Finally, fossil field properties do not generally appear to be strongly correlated with basic stellar properties such as mass, temperature or rotation rate \citep[see e.g. the reviews of][]{landstreet1982,mathys2001,donati2009}.

For the last 10-15 years, various studies have improved our knowledge of the fossil field properties of intermediate- and high-mass stars. The high-resolution spectropolarimetric survey of about 70 Herbig Ae/Be stars (HAeBes) has demonstrated that $\sim$7\% of Herbig Ae/Be stars host large-scale dipolar fields stronger than $\sim$300~G, very similar to the fields observed on the MS. Herbig Ae/Be stars are still contracting on the pre-main-sequence (PMS). Assuming the conservation of magnetic flux, the observed strengths of their fields are compatible with the bulk strengths of the magnetic fields of the main sequence A/B stars. It therefore appears that fossil fields are already present on the PMS \citep[e.g.][]{alecian2009,alecian2013}.

The Magnetism in Massive Stars (MiMeS) and B-fields in OB stars (BOB) projects, for the first time, have brought stringent high-resolution spectropolarimetric constraints on the magnetic fields of OB stars (spectral types B3 and earlier). Fossil fields also appear to be present in about 7\% of the more than 500 observed OB stars \citep{grunhut2012c,fossati2015b,wade2016,scholler2017,grunhut2017}. In general, these fields have topologies and strengths similar to the fossil fields of A/B stars \citep{wade2012,grunhut2012a,neiner2012,grunhut2013,wade2015,oksala2015}. However, a few stars display magnetic fields weaker than 300~G, suggesting that the lower limit of 300~G observed in A/B stars may be lower in massive stars. The magnetic dichotomy may therefore depend on stellar mass, with a magnetic gap decreasing as stellar mass increases \citep{fossati2015a}.

A recent high-resolution spectropolarimetric survey in about 50 cool (G-K) red giants has brought interesting results on the magnetic fields of intermediate-mass stars at the post-MS phases of stellar evolution. Magnetic fields have been detected in 30 stars of their sample, at different evolutionary stages: from the beginning of the Hertzsprung gap up to the beginning of the AGB phase. Most of the detected magnetic fields display complex configurations and rapid evolution, which is typical of dynamo processes occurring in their extended convective envelope. However, four of them display peculiar magnetic properties compared to the other magnetic stars of their sample. These fields are much stronger than the fields expected at their rotation rates based on the larger magnetic sample, they are of simple configurations (dipolar), and are relatively stable on timescales of years \citep{auriere2008,auriere2011,auriere2012,auriere2015}. They are therefore presumably the descendents of magnetic A/B stars hosting fossil fields, suggesting that fossil fields may well survive beyond the MS.
Recently, the first few evolved hotter magnetic stars have also been discovered and show weak magnetic signatures compatible with dipolar fossil fields and magnetic flux conservation \citep{neiner2017}.

While the global topologies of fossil fields seem to be conserved from the PMS to the post-MS phases, the magnetic strengths observed at the surfaces of  stars seem to decay - at least statistically - with time. This has been observed on the MS in A/B and OB stars. The magnitude of this decay appears to be larger than that expected due to conservation of magnetic flux as the stars inflate during their MS evolution. It therefore seems that on the timescale of the MS lifetime, some magnetic flux is lost by A, B and O stars. It is not yet clear what is the origin of this loss. It is however relatively small, in the sense that even close to the TAMS, stars can still have strong magnetic fields \citep{landstreet2007,landstreet2008,fossati2016}.

Finally, recent studies using the four Stokes parameters ($IQUV$) in combination with tomographic mapping techniques (Zeeman Doppler Imaging) have allowed the analysis of both the line-of-sight and transverse components of the magnetic field at the surfaces of stars, providing us with more detailed and complete magnetic maps of A/B stars. It has been found, in several A/B stars, that departures from typical large-scale multipolar structures exist. The magnetic field of one early B-type star, HD 37776, has been shown to have a dominant non-axisymmetric topology, clearly showing a much more complex field. However, these fields are still stable over many years, which does not call into question their fossil nature. Although the number of such mapping studies of that kind is not yet large enough for definitive conclusions, it has been proposed that the complex component of fossil fields may be age and/or mass dependent \citep{silvester2014,silvester2015,rusomarov2015,rusomarov2016}.

\section{Towards an understanding of the origin of fossil fields}

The origin of these fields is still the subject of intense debate \citep[e.g.][]{neiner2015}. Their fossil nature, the absence of extended convective envelopes in OBA stars, as well as the detection of magnetic fields in totally radiative stars among the Herbig Ae/Be sample, do not favour an in situ magnetic generation. The fields have most likely been shaped during the formation of the stars and have not evolved much since \citep[see also][]{moss2001}. The challenge today is to understand how and at which stage they have been shaped. In particular it is a mystery why such a small fraction of ABO stars contains fossil fields, while roughly 90\% of them are not strongly magnetic. Various hypotheses on the possible origin of these fossil fields have been proposed. Some of them have been, or are being, tested, thanks to different observational projects. We describe some of them below.

\subsection{Molecular cloud fossil fields}

In the most simple view of the fossil field theory we assume that strong magnetic fields observed in the radiative envelopes of ABO stars are direct remnants of the Galactic magnetic fields threading the molecular clouds where stars form. During the molecular cloud contraction, magnetic fields are swept up and concentrated into the forming protostars. Direct and indirect evidence of magnetic fields are observed in protostars at various evolutionary stages. The strengths of the magnetic fields measured in molecular clouds are strong enough (in fact probably too strong) to produce kG-strength fields via flux conservation in stellar envelopes on the MS \citep{moss2001,mestel2001}, making  this simple hypothesis essentially plausible. However we know today that material from molecular clouds, before becoming part of a star, must experience various phases of contraction most likely via an accretion/turbulent disk, to ultimately form a convective and fast-rotating protostar, and continuing its contraction with the proto-star itself. It is therefore not clear how the initial magnetic flux will evolve - or even survive - all of these phases. It will most likely not stay intact. A possibility is that the evolution of the magnetic field may - or may not - influence the different phases for star formation to finally produce - or not - two categories of stars: magnetic and non-magnetic. In summary, the question we ask is, are the initial conditions of the parental molecular clouds responsible for the magnetic properties observed in ABO stars, and especially for the magnetic dichotomy ?

An efficient way to address this question is to study close binary systems, close enough so we can reasonably assume that both stars have been formed together with similar initial conditions. Measuring and analysing magnetic fields in the two components of such systems can help us to disentangle initial condition effects from other factors (e.g. evolution, rotation). In a broader context, the Binarity and Magnetic Interaction in various classes of Stars \citep[BinaMIcS,][]{alecian2015} project aims at determining the global properties of the magnetic fields in short-period double-lined spectroscopic binary (SB2) systems. The BinaMIcS sample has been divided into 2 components: one containing convective cool systems, and the other containing hot systems with radiative envelopes. The latter sample is the only one that is relevant for addressing the issue of fossil fields observable at stellar surfaces. The BinaMIcS hot-sample contains about 170 SB2s with both components being on the MS and of spectral type earlier than F5 (securing outer radiative envelopes), and with orbital periods below 20\,d. The distance between the two components of the most massive SB2 systems of this sample ($\sim$80\,\msun\ of total mass) is less than 1\,au for an orbital period of 20\,d. It is even smaller for less massive systems and at shorter periods. It is not well understood how such close systems form, but the most likely hypothesis is that both stars have been formed together from the same environment. Although stellar capture represents a possible formation mechanism in rare cases, it is not likely to be the main formation process of short-period SB2s. It is therefore reasonable to assume that the two stars of such systems have been formed in similar initial conditions at the same time.

We have obtained one or several high-resolution spectropolarimetric observations with the spectropolarimeters Narval and ESPaDOnS of each system in order to evaluate the presence of a magnetic field. If a magnetic field is detected, additional observations have been obtained to derive magnetic maps, and determine the field strengths and topologies, which can then be compared to those of the isolated ABO stars. Among the 170 systems, 5 magnetic fields have been detected, including 3 that were previously known \citep[HD 98088, HD 5550, $\epsilon$\,Lup\,Aa ;][]{babcock1958, carrier2002, shultz2015}, and 2 new detections. One of them has first been obtained in the secondary component of HD\,160922, an F4+F5 system \citep[][Neiner et al. in prep.]{neiner2013}. The magnetic secondary is at the limit of the theoretical convective/radiative envelope limit, and it is not yet clear if the detected magnetic field is dynamo- or fossil-like. The second new detection has been obtained in the second component of a previously known magnetic SB2, $\epsilon$\,Lup\,Ab, implying that both B-type components of the $\epsilon$\,Lup system host fossil fields \citep{shultz2015}.

\begin{figure}
\centering
\begin{minipage}[t]{.39\textwidth}
\centering
\includegraphics[height=6.1cm]{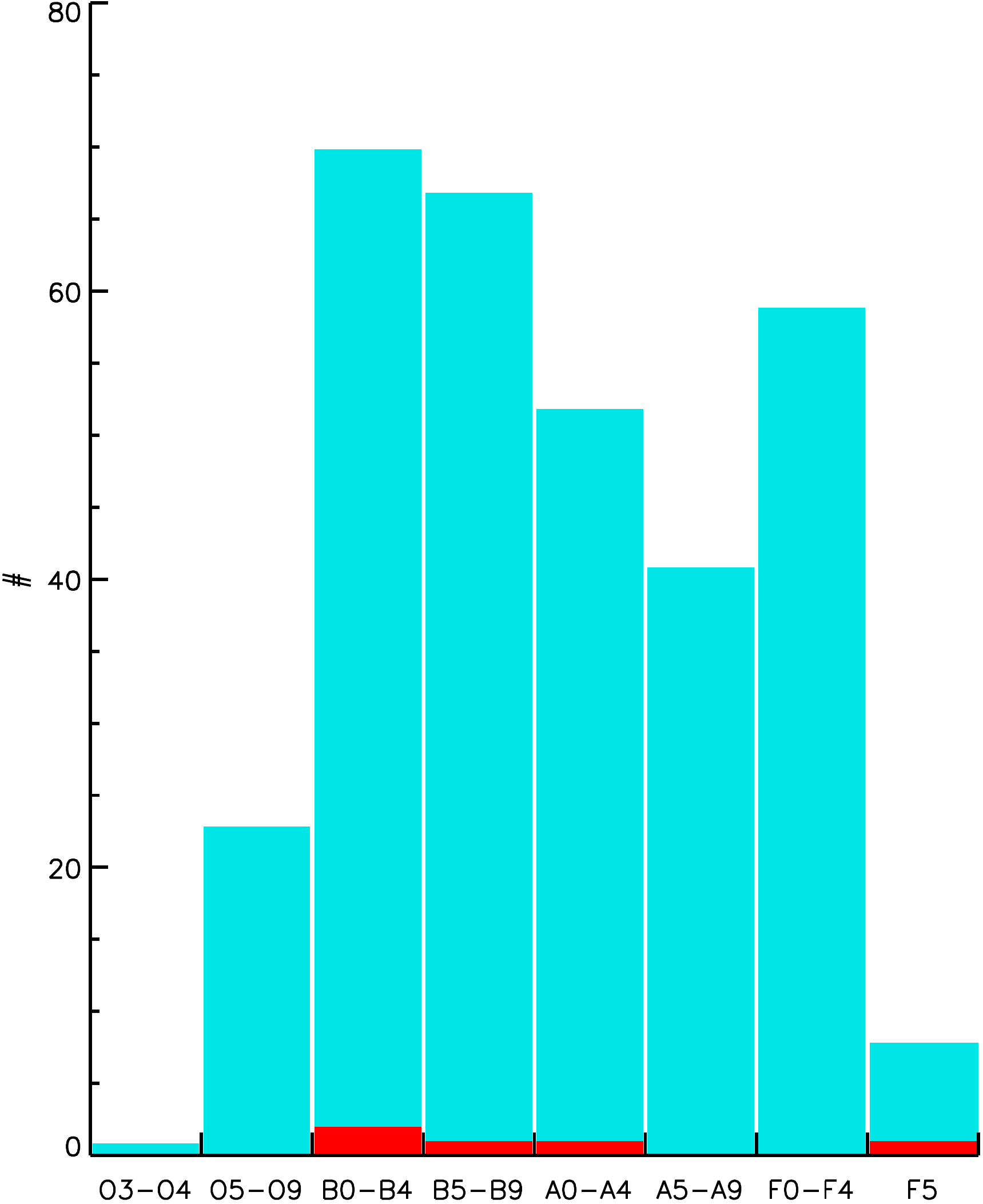}
\caption{Spectral type distribution of the components of short-period SB2 systems observed within the BinaMIcS project. {\it Light blue:} The whole sample. {\it Red:} Magnetic stars. (Alecian et al. in prep.)}
\label{fig:spt}
\end{minipage}
\hfill
\begin{minipage}[t]{.58\textwidth}
\centering
\includegraphics[height=6.0cm]{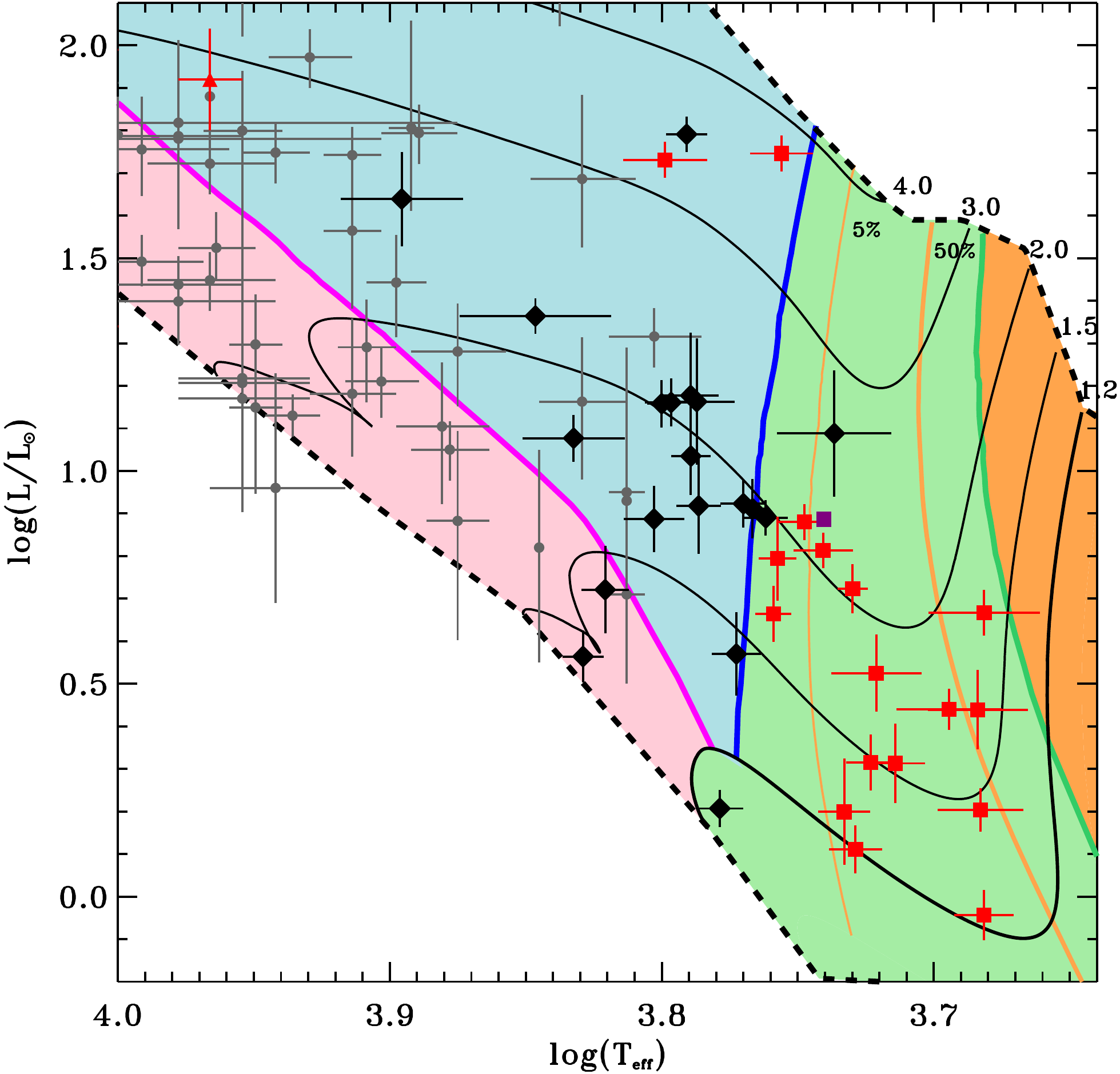}
\caption{HR diagram of the PMS phase of intermediate-mass stars. {\it Orange:} Fully convective. {\it Green:} Convective envelope + radiative core. {\it Blue:} Fully radiative. {\it Pink:} Radiative envelope + convective core. {\it Black full line:} PMS evolutionary tracks for 1.2, 1.5, 2, 3, and 4~\msun. {\it Black dashed lines}: The birthline and the ZAMS. {\it Orange full lines:} The convective envelope contains 50\% and 5\% of the stellar mass. {\it Red symbols:} Magnetic stars. {\it Black symbols:} Non-magnetic stars. {\it Small symbols:} HAeBes. {\it Large symbols:} Intermediate-mass T Tauri stars (IMTTS). (Villebrun et al. in prep.)}
\label{fig:hr}
\end{minipage}
\end{figure}

The analysis of the data is still in progress. However, magnetic configurations have already been obtained for 2 systems (HD 98088 and HD 5550) out of the 5 magnetic stars, and they show global properties similar to those of isolated ABO stars \citep{folsom2013,alecian2016}. The statistical analysis of the whole sample is also in progress, but we can already say that the magnetic incidence is much lower than in isolated stars: less than 1.5\% (versus 7\% in single stars, see Fig.\,\ref{fig:spt}). Furthermore, except in one system, fossil magnetic fields are present in only one of the two components. If initial conditions were solely responsible for the presence or not of a magnetic field in OBA, then we would naively expect that both components of the system should host fossil fields, which is not the case. This supports that the magnetic properties of ABO stars are more likely a product of the entire formation history of a star, rather than just the initial conditions of formation (Alecian et al. in prep.).

On the contrary, the lower incidence of fossil fields in short-period binaries as compared to isolated stars is a puzzling result. It is evident that both kinds of stars (isolated on one side, and with a close companion on the other side) do not have the same formation history. A better understanding of the formation of close binaries would help us in our understanding. In the meantime we can recall theoretical and modelling work performed by e.g. \citet{commercon2010} showing that in the presence of strong initial magnetic fields in a molecular cloud, fragmentation is reduced, making it more difficult to form binaries. The formation of binaries must involve fragmentation at some stage \citep[see also][]{forgan2017}. We can therefore wonder whether the initial reservoir of magnetic energy was larger in the case of isolated stars, than in the case of short-period binaries.

\subsection{Binary interaction}

It has been proposed by e.g. \cite{ferrario2009} that the fossil magnetic fields of single stars could be the result of merging processes. Ferrario et al. argue that the small fraction of magnetic stars could be due to a small fraction of merging processes occurring during the late phase of the PMS, once the stars have become totally radiative. The merger scenario predicts strong differential rotation that would generate a stable and large-scale magnetic field. If this hypothesis is true, all (or at least a substantial number) of merger products should host strong and stable magnetic fields. Such an hypothesis could potentially be tested by studying blue stragglers.

Blue stragglers are MS stars and highly probable members of open or globular clusters, but appear more luminous and hotter than the turn-off point of the cluster in the HR diagram. It has been proposed they may be the results of merging of two stars \citep[][]{mccrea1964,hills1976}. If such an interaction is at the origin of fossil fields, basically all of blue stragglers would host fossil magnetic fields. We have obtained spectropolarimetric observations of 33 blue stragglers, and detected magnetic fields only in about 6\% of them - a fraction similar to the incidence in isolated MS massive stars (Grunhut et al. 2017).

In addition, the detection of fossil magnetic fields in close binaries, and in particular in the two components of the short-period binary $\epsilon$\,Lup, makes the merger scenario even less likely as it is difficult to understand how merging could have occurred in the past in such close systems. We therefore tentatively conclude that merging does not appear to be the main process at the origin of fossil fields.

It has also been proposed that strong shear under the surface of the components of 'almost-merging' systems, experiencing, or having experienced in the past, mass-transfer episode or a common-envelope phase, could be at the origin of fossil fields in O-type stars \citep[e.g.][]{langer2014}. The first candidate is Plaskett's star in which a fossil-like magnetic field has been detected in the secondary component \citep{grunhut2013}. Following its detection, \cite{naze2017} have suggested that Plaskett-like stars (so-called 'Plaskett's twins'), i.e. interacting or post-interacting massive binaries, may all become magnetic thanks to mass transfer processes. They compiled a sample of 15 such systems and analyzed spectropolarimetric measurements from the literature and from the BinaMIcS project. They detected no magnetic fields. By taking into account the magnetic sensitivity of the data, they concluded that in Plaskett-like objects, the magnetic incidence is similar to that found by the larger MiMeS and BOB surveys of isolated stars. Therefore mass transfer does not appear to play a dominant role in the origin of fossil fields either.

\subsection{Relaxed magnetic fields}

It is now generally agreed that stars at all masses form from a fully convective low-mass core, which grows through accretion during the protostellar phase \citep[e.g.][]{palla1992,behrend2001}. Stars rise upwards along the birthline in the HR diagram, growing in mass until the strong accretion phase stops, then continue to evolve with a quasi-static contraction along PMS evolutionary tracks (see Fig.\,\ref{fig:hr}). At intermediate and high masses (above 1.5\,M$_{\odot}$), during their evolution on the birthline and on the PMS tracks, the interiors of the stars experience profound changes. They first start fully convective, then a radiative core forms and grows until it reaches the surface of the star. At that point the stars are fully radiative until a small convective core develops right before they reach the ZAMS. Such structural changes have important implications for the properties and evolution of surface magnetic fields. During the convective phase, where the stars are either fully convective or have a convective envelope (orange and green regions in Fig.\,\ref{fig:hr}), complex and/or variable magnetic fields are generated or strengthened via a dynamo process. In the radiative phases (the blue and pink regions in Fig.\,\ref{fig:hr}), no strong dynamo fields are generated in the envelope, while the dynamo field of the core takes too long to appear at the surface \citep[e.g.][]{moss2001}. The limit between the two regimes (the thick blue line in Fig.\,\ref{fig:hr}) is called the convective/radiative (CR) limit and constitutes the part of the HR diagram that we will focus on in the following.

The relaxed magnetic field theory considers that fossil fields are remnants of fields generated during the convective phase of the PMS stellar evolution. The basic idea is that, for all stars, during the transition from convective to radiative, magnetic fields present in the convective zone relax into the radiative zone. The relaxed fields can have various configurations. Only 7\% of them will end up with magnetic fields with a mixed (poloidal and toroidal) configuration, which are therefore stable, and can last for the lifetime of a star. Theoretical and numerical investigations have brought strong arguments in favour of this "basic idea". In the non-rotating case, numerical models have demonstrated that fields generated in a previous convective phase can relax into fields of mixed configuration during the transition from the convective to the radiative phase \citep{duez2010a,duez2010b,duez2011,arlt2014}. Recent theoretical developments have shown that, in the case of very high rotation rates, the relaxed fields have a strong toroidal component, that may violate the conditions for stability \citep{emeriau2015}. Rotation appears therefore to play a major role in fossil field formation. Those works have also shown that helicity plays an important role in the final configuration of the relaxed fields.

We have recently started an observing program to characterize the magnetic fields of intermediate-mass T Tauri stars (IMTTS), i.e. stars covering the convective zones of the HR diagram at intermediate-mass, as well as the CR limit. These stars are the progenitors of the Herbig Ae/Be stars. The objectives of this project is to understand the evolution of the rotation and magnetic fields along the PMS convective phases until the stars become totally radiative, and therefore to understand the emergence of the fossil fields, and the conditions under which they emerge.

We have first performed snapshot observations of about 50 IMTTS to detect their magnetic fields. We have detected magnetic fields in about 50\% of our sample, which contrasts with the magnetic incidence in low-mass T Tauri stars, close to 100\%, and with the ~7\% magnetic incidence in HAeBes. To understand those differences, we have performed careful determination of the effective temperatures and luminosities, using GAIA distances whenever available, to place the stars in the HR diagram. The result is shown in Fig.\,\ref{fig:hr}. We observe that in the convective regions of the HR diagram the magnetic detection rate is close to 100\%, while stars in which magnetic fields have not been detected are situated in the radiative regions. We observe also that very quickly, once the stars have crossed the CR limit, we reach an incidence rate close to 6\%, similar to the incidence rates of fossil fields measured in the Herbig and MS stars (Villebrun et al. in prep.). These first results are in favour of the relaxed field theory. We are currently acquiring additional data to determine the topologies of the magnetic fields of part of our sample at different stages of the convective phase, i.e. in stars with different masses of the convective envelope, in order to bring observational constraints to the relaxed field theory.

\section{Discussion}


Today, the relaxed field theory is the most promising one for explaining the origin of fossil fields in intermediate- and high-mass stars. The initial conditions of the parental molecular cloud, in which stars form, may have an impact on the magnetic properties observed at the MS stage, but are most likely not dominant. The entire formation history must be considered. In particular, we must also take into account the interaction of the PMS stars with their environment : accretion, outflows, disk interaction, tidal interaction. These processes may strongly affect, depending on the evolutionary stage, the magnetic flux and angular momentum distribution at the surface of the star, hence inside the star. Differential rotation can indeed lead to instabilities strong enough to dissipate large-scale magnetic fields \citep{gaurat2015}. The study of the sample of IMTTS partly presented above is a first step to understand how the convective phase can impact fossil field properties. It is also aimed at analysing the rotation and accretion properties of the sample, to understand their interplay with the magnetic fields. On a longer timescale, the near-IR high-resolution spectropolarimeter SPIRou, as well as similar instruments that will be on-sky in the coming year, will allow us to characterise magnetic fields and rotation of the class I progenitors of the IMTTS, as well as the magnetic fields of the disks of class I  and class II objects, providing us with additional observational constraints to aid in our understanding of the origin of fossil magnetic fields.


\section*{Acknowledgement}

We acknowledge financial support from PNPS of CNRS/INSU, France, and from the AGIR program of the "Universit\'e Grenoble Alpes". This work has made use of the SIMBAD database, VizieR catalogue access tool, "Aladin sky atlas", CDS, Strasbourg, France, and of data from the ESA mission {\it Gaia}, processed by the {\it Gaia} Data Processing and Analysis Consortium (DPAC, \url{https://www.cosmos.esa.int/web/gaia/dpac/consortium}). GAW acknowledges Discovery Grant support from the Natural Sciences and Engineering Research Council (NSERC) of Canada. 

\bibliography{mnemonic_ea,alecian_astrofluid2016}

\begin{thebibliography}{}

\bibitem[\protect\astroncite{Alecian et~al.}{2015}]{alecian2015}
Alecian E., Neiner C., et~al.: 2015,
\newblock {\em New windows on massive stars: asteroseismology} {\bf 307}, 330

\bibitem[\protect\astroncite{Alecian et~al.}{2016}]{alecian2016}
Alecian E., Tkachenko A., et~al.: 2016,
\newblock {\em A\&A} {\bf 589}, A47

\bibitem[\protect\astroncite{Alecian et~al.}{2009}]{alecian2009}
Alecian E., Wade G.~A., et~al.: 2009,
\newblock {\em MNRAS} {\bf 400}, 354

\bibitem[\protect\astroncite{Alecian et~al.}{2013}]{alecian2013}
Alecian E., Wade G.~A., et~al.: 2013,
\newblock {\em MNRAS} {\bf 429}, 1001

\bibitem[\protect\astroncite{Arlt}{2014}]{arlt2014}
Arlt R.: 2014,
\newblock {\em Putting A Stars into Context: Evolution} pp 93--101

\bibitem[\protect\astroncite{Auri{\`e}re et~al.}{2015}]{auriere2015}
Auri{\`e}re M., Konstantinova-Antova R., et~al.: 2015,
\newblock {\em A\&A} {\bf 574}, A90

\bibitem[\protect\astroncite{Auri{\`e}re et~al.}{2008}]{auriere2008}
Auri{\`e}re M., Konstantinova-Antova R., et~al.: 2008,
\newblock {\em A\&A} {\bf 491}, 499

\bibitem[\protect\astroncite{Auri{\`e}re et~al.}{2012}]{auriere2012}
Auri{\`e}re M., Konstantinova-Antova R., et~al.: 2012,
\newblock {\em A\&A} {\bf 543}, A118

\bibitem[\protect\astroncite{Auri{\`e}re et~al.}{2011}]{auriere2011}
Auri{\`e}re M., Konstantinova-Antova R., et~al.: 2011,
\newblock {\em A\&A} {\bf 534}, A139

\bibitem[\protect\astroncite{Auri{\`e}re et~al.}{2007}]{auriere2007}
Auri{\`e}re M., Wade G.~A., et~al.: 2007,
\newblock {\em A\&A} {\bf 475}, 1053

\bibitem[\protect\astroncite{Babcock}{1958}]{babcock1958}
Babcock H.~W.: 1958,
\newblock {\em ApJS} {\bf 3}, 141

\bibitem[\protect\astroncite{Behrend and Maeder}{2001}]{behrend2001}
Behrend R. and Maeder A.: 2001,
\newblock {\em A\&A} {\bf 373}, 190

\bibitem[\protect\astroncite{Blaz{\`e}re et~al.}{2016b}]{blazere2016b}
Blaz{\`e}re A., Neiner C., and Petit P.: 2016b,
\newblock {\em MNRAS} {\bf 459}, L81

\bibitem[\protect\astroncite{Blaz{\`e}re et~al.}{2016a}]{blazere2016a}
Blaz{\`e}re A., Petit P., et~al.: 2016a,
\newblock {\em A\&A} {\bf 586}, A97

\bibitem[\protect\astroncite{Borra et~al.}{1982}]{borra1982}
Borra E.~F., Landstreet J.~D., and Mestel L.: 1982,
\newblock {\em ARA\&A} {\bf 20}, 191

\bibitem[\protect\astroncite{Carrier et~al.}{2002}]{carrier2002}
Carrier F., North P., et~al.: 2002,
\newblock {\em A\&A} {\bf 394}, 151

\bibitem[\protect\astroncite{Commercon et~al.}{2010}]{commercon2010}
Commercon B., Hennebelle P., et~al.: 2010,
\newblock {\em A\&A} {\bf 510}, L3

\bibitem[\protect\astroncite{Donati and Landstreet}{2009}]{donati2009}
Donati J.~F. and Landstreet J.~D.: 2009,
\newblock {\em ARA\&A} {\bf 47}, 333

\bibitem[\protect\astroncite{Duez}{2011}]{duez2011}
Duez V.: 2011,
\newblock {\em Astron. Nachr.} {\bf 332}, 983

\bibitem[\protect\astroncite{Duez et~al.}{2010}]{duez2010b}
Duez V., Braithwaite J., and Mathis S.: {2010},
\newblock {\em ApJ} {\bf 724}, L34

\bibitem[\protect\astroncite{Duez and Mathis}{2010}]{duez2010a}
Duez V. and Mathis S.: {2010},
\newblock {\em A\&A} {\bf 517}, A58

\bibitem[\protect\astroncite{Emeriau and Mathis}{2015}]{emeriau2015}
Emeriau C. and Mathis S.: 2015,
\newblock {\em New wind. on mass. stars: asteroseis.} {\bf 307}, 373

\bibitem[\protect\astroncite{Ferrario et~al.}{2009}]{ferrario2009}
Ferrario L., Pringle J.~E., et~al.: 2009,
\newblock {\em MNRAS} {\bf 400}, L71

\bibitem[\protect\astroncite{Folsom et~al.}{2013}]{folsom2013}
Folsom C.~P., Likuski K., et~al.: 2013,
\newblock {\em MNRAS} {\bf 431}, 1513

\bibitem[\protect\astroncite{Forgan et~al.}{2017}]{forgan2017}
Forgan D., Price D.~J., and Bonnell I.: 2017,
\newblock {\em MNRAS} {\bf 466}, 3406

\bibitem[\protect\astroncite{Fossati et~al.}{2015a}]{fossati2015a}
Fossati L., Castro N., et~al.: {2015a},
\newblock {\em A\&A} {\bf 574}, A20

\bibitem[\protect\astroncite{Fossati et~al.}{2015}]{fossati2015b}
Fossati L., Castro N., et~al.: 2015,
\newblock {\em A\&A} {\bf 582}, A45

\bibitem[\protect\astroncite{Fossati et~al.}{2016}]{fossati2016}
Fossati L., Schneider F. R.~N., et~al.: 2016,
\newblock {\em A\&A} {\bf 592}, A84

\bibitem[\protect\astroncite{Gaurat et~al.}{2015}]{gaurat2015}
Gaurat M., Jouve L., et~al.: 2015,
\newblock {\em A\&A} {\bf 580}, A103

\bibitem[\protect\astroncite{Grunhut et~al.}{2012a}]{grunhut2012a}
Grunhut J.~H., Rivinius T., et~al.: {2012a},
\newblock {\em MNRAS} {\bf 419}, 1610

\bibitem[\protect\astroncite{Grunhut et~al.}{2013}]{grunhut2013}
Grunhut J.~H., Wade G.~A., et~al.: 2013,
\newblock {\em MNRAS} {\bf 428}, 1686

\bibitem[\protect\astroncite{Grunhut et~al.}{2017}]{grunhut2017}
Grunhut J.~H., Wade G.~A., et~al.: 2017,
\newblock {\em MNRAS} {\bf 465}, 2432

\bibitem[\protect\astroncite{Grunhut et~al.}{2012c}]{grunhut2012c}
Grunhut J.~H., Wade G.~A., and {the MiMeS Collaboration}: {2012c},
\newblock in {\em Proceedings of a Scientific Meeting in Honor of Anthony F. J.
  Moffat}, pp 42--

\bibitem[\protect\astroncite{Hills and Day}{1976}]{hills1976}
Hills J.~G. and Day C.~A.: 1976,
\newblock {\em Astrophys. Lett.} {\bf 17}, 87

\bibitem[\protect\astroncite{Landstreet}{1982}]{landstreet1982}
Landstreet J.~D.: 1982,
\newblock {\em ApJ} {\bf 258}, 639

\bibitem[\protect\astroncite{Landstreet et~al.}{2007}]{landstreet2007}
Landstreet J.~D., Bagnulo S., et~al.: 2007,
\newblock {\em A\&A} {\bf 470}, 685

\bibitem[\protect\astroncite{Landstreet et~al.}{2008}]{landstreet2008}
Landstreet J.~D., Silaj J., et~al.: 2008,
\newblock {\em A\&A} {\bf 481}, 465

\bibitem[\protect\astroncite{Langer}{2014}]{langer2014}
Langer N.: 2014,
\newblock {\em IAU Symposium} {\bf 302}, 1

\bibitem[\protect\astroncite{Mathys}{2001}]{mathys2001}
Mathys G.: 2001,
\newblock {\em Magn. Fields Across the Hertzsprung-Russell Diagram} {\bf 248},
  267

\bibitem[\protect\astroncite{McCrea}{1964}]{mccrea1964}
McCrea W.~H.: 1964,
\newblock {\em MNRAS} {\bf 128}, 147

\bibitem[\protect\astroncite{Mestel}{2001}]{mestel2001}
Mestel L.: 2001,
\newblock {\em Magnetic Fields Across the Hertzsprung-Russell Diagram} {\bf
  248}, 3

\bibitem[\protect\astroncite{Moss}{2001}]{moss2001}
Moss D.: 2001,
\newblock {\em Magn. Fields Across the Hertzsprung-Russell Diagram} {\bf 248},
  305

\bibitem[\protect\astroncite{Naz{\'e} et~al.}{2017}]{naze2017}
Naz{\'e} Y., Neiner C., et~al.: 2017,
\newblock {\em MNRAS} {\bf 467}, 501

\bibitem[\protect\astroncite{Neiner and Alecian}{2013}]{neiner2013}
Neiner C. and Alecian E.: 2013,
\newblock {\em EAS Publications Series} {\bf 64}, 75

\bibitem[\protect\astroncite{Neiner et~al.}{2012}]{neiner2012}
Neiner C., Alecian E., et~al.: 2012,
\newblock {\em A\&A} {\bf 537}, A148

\bibitem[\protect\astroncite{{Neiner} et~al.}{2015}]{neiner2015}
{Neiner} C., {Mathis} S., et~al.: 2015,
\newblock {\em IAU Symposium} {\bf 305}, 61

\bibitem[\protect\astroncite{Neiner et~al.}{2017}]{neiner2017}
Neiner C., Oksala M., et~al.: 2017,
\newblock {\em A\&A} in press

\bibitem[\protect\astroncite{Oksala et~al.}{2015}]{oksala2015}
Oksala M.~E., Kochukhov O., et~al.: 2015,
\newblock {\em MNRAS} {\bf 451}, 2015

\bibitem[\protect\astroncite{Palla and Stahler}{1992}]{palla1992}
Palla F. and Stahler S.~W.: 1992,
\newblock {\em ApJ} {\bf 392}, 667

\bibitem[\protect\astroncite{Petit et~al.}{2011}]{petit2011}
Petit P., Ligni{\`e}res F., et~al.: 2011,
\newblock {\em A\&A} {\bf 532}, L13

\bibitem[\protect\astroncite{Rusomarov et~al.}{2016}]{rusomarov2016}
Rusomarov N., Kochukhov O., et~al.: 2016,
\newblock {\em A\&A} {\bf 588}, A138

\bibitem[\protect\astroncite{Rusomarov et~al.}{2015}]{rusomarov2015}
Rusomarov N., Kochukhov O., et~al.: 2015,
\newblock {\em A\&A} {\bf 573}, A123

\bibitem[\protect\astroncite{Sch{\"o}ller et~al.}{2017}]{scholler2017}
Sch{\"o}ller M., Hubrig S., et~al.: 2017,
\newblock {\em A\&A} {\bf 599}, A66

\bibitem[\protect\astroncite{Shultz et~al.}{2015}]{shultz2015}
Shultz M., Wade G.~A., et~al.: 2015,
\newblock {\em MNRAS} {\bf 454}, L1

\bibitem[\protect\astroncite{Silvester et~al.}{2014}]{silvester2014}
Silvester J., Kochukhov O., and Wade G.~A.: 2014,
\newblock {\em MNRAS} {\bf 440}, 182

\bibitem[\protect\astroncite{Silvester et~al.}{2015}]{silvester2015}
Silvester J., Kochukhov O., and Wade G.~A.: 2015,
\newblock {\em MNRAS} {\bf 453}, 2163

\bibitem[\protect\astroncite{Wade et~al.}{2015}]{wade2015}
Wade G.~A., Barb{\'a} R.~H., et~al.: 2015,
\newblock {\em MNRAS} {\bf 447}, 2551

\bibitem[\protect\astroncite{Wade et~al.}{2012}]{wade2012}
Wade G.~A., Ma{\'\i}z~Apell{\'a}niz J., et~al.: 2012,
\newblock {\em MNRAS} {\bf 425}, 1278

\bibitem[\protect\astroncite{Wade et~al.}{2016}]{wade2016}
Wade G.~A., Neiner C., et~al.: 2016,
\newblock {\em MNRAS} {\bf 456}, 2

\end{thebibliography}

\end{document}